\documentclass[structabstract, ]{aa} 

\usepackage{natbib}
\bibpunct{(}{)}{;}{a}{}{,} 
\usepackage{graphicx, amsmath}
\usepackage{txfonts}

\newcommand{\zxipcf}{{\fontfamily{ptm}\fontseries{m}\fontshape{sc}\selectfont{zxipcf}}}
\newcommand{\tbabs}{{\fontfamily{ptm}\fontseries{m}\fontshape{sc}\selectfont{tbabs}}}
\newcommand{\ztbabs}{{\fontfamily{ptm}\fontseries{m}\fontshape{sc}\selectfont{ztbabs}}}
\newcommand{\pexmon}{{\fontfamily{ptm}\fontseries{m}\fontshape{sc}\selectfont{pexmon}}}
\newcommand{\pexrav}{{\fontfamily{ptm}\fontseries{m}\fontshape{sc}\selectfont{pexrav}}}

\newcommand{\xstar}{{\fontfamily{ptm}\fontseries{m}\fontshape{sc}\selectfont{xstar}}}

\newcommand{\source}{{Fairall 51}}

\newcommand{\xmm}{{{XMM-Newton}}}

\newcommand{\swift}{{Swift}}

\newcommand{\suzaku}{{Suzaku}}

\newcommand{\xis}{{XIS}}
\newcommand{\hxd}{{HXD/PIN}}

\begin{document}
\title{An X-ray variable absorber within the Broad Line Region in Fairall~51}
\author{J.~Svoboda\inst{\ref{inst1}}\thanks{email: jiri.svoboda@asu.cas.cz}
\and T.~Beuchert\inst{\ref{inst2}}
\and M.~Guainazzi\inst{\ref{inst3}}
\and A.~L.~Longinotti\inst{\ref{inst3},\ref{inst4}}
\and E.~Piconcelli\inst{\ref{inst5}}
\and J.~Wilms\inst{\ref{inst2}}
}
\institute{Astronomical Institute, Academy of Sciences, Bo\v{c}n\'{\i}~II~1401, CZ-14100~Prague, Czech~Republic\label{inst1}
\and Dr. Karl Remeis Sternwarte, Sternwartstrasse 7, D-96049 Bamberg, Germany\label{inst2}
\and European Space Astronomy Centre of ESA, PO Box 78, Villanueva de la Ca\~{n}ada, 28691 Madrid, Spain\label{inst3}
\and Departamento de Astrofisica Extragalactica y Cosmologia, Instituto de Astronomia, Universidad Nacional Autonoma de Mexico (UNAM), Apartado Postal 70-264, 04510 Mexico\label{inst4}
\and Osservatorio Astronomico di Roma (INAF), via Frascati 33, I-00040 Monteporzio Catone Roma, Italy\label{inst5}
}


\abstract{Fairall 51 is a polar-scattered Seyfert 1 galaxy,
a type of active galaxies believed 
to represent a bridge between unobscured type-1 and obscured type-2 objects. 
Fairall 51 has shown complex and variable X-ray absorption
but only little is known about its origin. }
{In our research, we observed Fairall 51 with the X-ray satellite Suzaku
in order to constrain a characteristic time-scale of its variability.}
{We performed timing and spectral analysis of four observations 
separated by 1.5, 2 and 5.5 day intervals.
}
{We found that the 0.5--50 keV broadband X-ray spectra are dominated 
by a primary power-law emission (with the photon index $\approx$ 2). 
This emission is affected by at least three absorbers with different 
ionisations ($\log\xi \approx 1$--$4$).
The spectrum is further shaped by a reprocessed emission, possibly
coming from two regions -- the accretion disc and a more distant scattering region.
The accretion disc emission is smeared by the relativistic effects,
from which we measured the spin of the black hole as $a \approx 0.8 \pm 0.2$.
We found that most of the spectral variability can be attributed 
to the least ionised absorber whose column density changed by a factor of two 
between the first (highest-flux) and the last (lowest-flux) observation. }
{A week-long scale of the variability indicates that the absorber is located 
at the distance $\approx 0.05$\,pc from the centre, i.e., in the
Broad Line Region.}



\maketitle

\section{Introduction}

In the unification paradigm of active galactic nuclei (AGN) by \citet{1993ARA&A..31..473A},
type~1 and 2 AGN are distinguished by the inclination angle,
more specifically whether 
or not the torus intercepts the line-of-sight.
However, this standard picture assuming 
a homogeneous dusty torus
has been questioned by several studies
\citep{2002ApJ...570L...9N, elitzur2012, Merloni2013}. 
The clumpiness of the absorber is supposed to be another important parameter
for the proper classification.
The type of AGN is not entirely
determined by the inclination angle but also by 
a probability
of the absorption clouds intercepting the line of sight.
This is supported by a large and fast variability
of the absorber observed in several sources 
\citep[see, e.g.,][]{2005ApJ...623L..93R, 2012AdAst2012E..17B}.
The statistical analysis based on the vast RXTE archive
was recently done by \citet{Markowitz2014}
giving the probability of an absorption event
regardless the constant absorption due to non-clumpy
material to be $0.006^{+0.160}_{-0.003}$ for type~1,
and $0.110^{+0.461}_{-0.071}$ for type~2 AGN, respectively.

The absorption variability occurs at different time scales
for different AGN.
\citet{2002ApJ...571..234R}
investigated a sample of Seyfert~2
galaxies 
and revealed that most of them are variable on time scales
of months to years.
Similar variability was recently reported based on {\suzaku} observations, e.g.,
by \citet{2013MNRAS.428.2516B} for NGC 4507, and
\citet{Rivers2014} for NGC 2110.
The variability time scale of months to years is typical 
for a clumpy torus at the distance $\approx$\,1\,pc.
However, the absorption variability was later found on the time-scale of days
in several sources: NGC 4388 \citep{2004ApJ...615L..25E}, 
NGC~1365 \citep{2005ApJ...623L..93R}, 
NGC~4151 \citep{2007MNRAS.377..607P}, 
NGC~7582 \citep{2009ApJ...695..781B}, 
PG~1535+547 \citep{2008A&A...483..137B}, 
Mrk~766 \citep{Risaliti2011},
NGC~5506 \citep{Markowitz2014},
NGC~3227 \citep{lamer2003, Beuchert2014},
or Swift J2127.4+5654 \citep{Sanfrutos2013}.
The origin of such variability must be significantly closer to the central
region.
Recently,
\citet{Torricelli2014} showed that X-ray eclipses are common when
the expected occultation time is compatible with the exposure
and that these events may be responsible for most of the spectral variability 
at energies higher than 2\,keV.
They associated the origin of the eclipsing clouds to the Broad Line Region of AGN.

Knowledge of the location of the absorber in the so called polar-scattered Seyfert~1 AGN 
is important to understand the nature of this kind of objects.
Generally, Seyfert~1 galaxies do not exhibit polarised emission.
It is believed that the total polarisation
is washed out by a mutual interaction of polarised emission
from two scattering
regions -- the equatorial plane and the ionisation cone \citep{2004MNRAS.350..140S, Batcheldor2011}. 
The equatorial emission is obscured by a dusty torus in Seyfert~2 galaxies
and thus, the polarised emission is detected only from the ionisation cone.
\citet{2004MNRAS.350..140S} argued that the detected polarisation is inclination-dependent
and that polar-scattered Seyfert~1 galaxies represent a bridge between
type 1 and type 2 galaxies.
In this scenario, the enhanced variability of the absorber (with respect to the equatorial-scattered ones)
is therefore expected because we may be seeing the AGN along a line-of-sight grazing the torus rim.

X-ray properties of polar-scattered Seyfert~1 galaxies were studied
by \citet{2008RMxAC..32..131J} 
who realised that the luminosity and the index of a power-law emission
are consistent with Seyfert~1 galaxies.
A low inclination is also suggested from detection
of a relativistically smeared iron line profile 
coming from the innermost accretion disc in most of the sources.
The variability and complex absorption are common features of these sources.

More recently, investigation of the variability properties
was done for several sources.
A clear example of the absorption variability
was shown in the case of Mrk 704 \citep{2011A&A...533A...1M, 2011ApJ...734...75L} 
where the spectra from two different observations separated by three years
were identical above 7\,keV but very different at lower energies.
The observed X-ray spectral variability of Mrk~231 below 10 keV 
can be interpreted as changes in a patchy absorber 
\citep{Piconcelli2013,Teng2014}.
\citet{Beuchert2014} found the presence of
a non-dusty absorber in NGC~3227 
varying on a time scale of  about one week.
They found that the absorber is intermediately ionised and partially
covering the source.
The spectral variability of another polar-scattered Seyfert~1 galaxy, ESO 323-G77, 
has been recently studied by \citet{Miniutti2014}.
They found two variable absorbers with a different
time-scale of their variability.
The absorbers were associated
to Broad-Line-Region clouds (month-long time scale) 
and a clumpy torus (year-long time scale) according to their variability.

{\source} (also known as {{ESO 140-43}}) is 
a nearby galaxy with the cosmological redshift
$z=0.0141$, as measured in the 2MASS survey \citep{Huchra2012}.
{\source} is classified as Seyfert~1 since its optical spectrum 
contains broad lines with the full width at half maximum  
$FWHM\approx 3000 \pm 1000$ km\,s$^{-1}$ \citep{Schmid2001}.
The optical/UV flux is highly polarised \citep{martin1983}.
VLT-spectropolarimetry measurements by \citet{Schmid2001}
revealed a polarisation degree ranging from $5\%$ (red) to $13\%$ (UV),
which are one of the highest polarisation degrees observed for type~1 objects.

{\source} is very bright in X-rays (about 1 millicrab).
It was observed twice with high-quality X-ray spectrometers
on-board the {\xmm} satellite. 
The first observation was performed in 2005 
with the observed flux $f^{\rm high}_{2-10\,{\rm keV}} \approx 3\times10^{-11}$\,erg\,cm$^{-2}$\,s$^{-1}$,
and the second one about half a year later with the observed flux
$f^{\rm low}_{2-10\,{\rm keV}} \approx 0.9\times10^{-11}$\,erg\,cm$^{-2}$\,s$^{-1}$.
The spectra significantly differ and
the complex soft X-ray spectrum may be described
by the presence of three zones of a partially-covering warm absorber
with high, moderate and very low ionisation \citep{ricci2010}. 

Short-time variability of this source was suggested from two {\swift}
observations performed in 2008. The observations were separated only by five days. 
Although the exposures were very short (a few ks each) the spectra
were qualitatively different below 7\,keV \citep{ricci2010, beuchert2013}.
\citet{beuchert2013} determined that the variable absorber must be located
not further than the Broad Line Region based on the short variability time-scale.

In this paper, we present the results of a recent monitoring of {\source}
by the Japanese X-ray satellite {\suzaku} \citep{2007PASJ...59S...1M},
during which four 30\,ks exposures were performed separated by 1.5, 2 and 5.5 days.
The paper is organised as follows:
the observations and the data reduction are described
in Sect.~\ref{reduction}. Results are presented in Sect.~\ref{results}
and discussed in Sect.~\ref{discussion}. Main conclusions are summarised
in Sect.~\ref{conclusion}.

\begin{table*}[tb]

\caption{List of {\suzaku} observations from September 2013.}
\centering
\begin{tabular}{ccccc}
\rule{0cm}{0.5cm}

Observation & Identification Number & Start Time & Stop Time & Net Exposure [ks]\\
\hline \hline
\rule[-0.7em]{0pt}{2em} 
1 & 708046010 & Sep 4th 01:45:56 & Sep 4th 12:31:08 & 31.5 \\
\rule[-0.7em]{0pt}{2em} 
2 & 708046020 & Sep 5th 19:22:38  & Sep 6th 03:59:00 & 31.0 \\
\rule[-0.7em]{0pt}{2em} 
3 & 708046030 & Sep 7th 19:35:09 & Sep 8th 10:17:13   & 24.4 \\
\rule[-0.7em]{0pt}{2em} 
4 & 708046040 & Sep 13th 12:16:30 & Sep 14th 07:32:52  & 30.4 \\
\end{tabular}

\label{list_observations} 
\end{table*}

\section{Observations and data reduction}
\label{reduction}

We observed {\source} using the X-ray {\suzaku} satellite \citep{2007PASJ...59S...1M} 
during four $\approx$30\,ks long exposures
in the first half of September 2013 (see Table~\ref{list_observations}). 
The observation identification numbers
are 708046010, 708046020, 708046030, and 708046040, respectively. All the observations
were performed at the XIS-nominal pointing position.

The Heasoft package version 6.14\footnote{http://heasarc.nasa.gov/lheasoft/}
was used for the data reduction and also for the subsequent spectral and timing analysis.
The data were processed standardly following the {{Suzaku Data
Reduction Guide}}\footnote{http://heasarc.nasa.gov/docs/suzaku/analysis/abc/} (version 4).
For all {\xis} detectors, we combined both 3x3 and 5x5 modes
to extract the event files.
The source spectra  were obtained from a circle 
around the centre of the point spread function
with the radius of $260''$.
We defined the background extraction region 
as an annulus around the source circle
with the outer radius of $360''$
to avoid any contamination from the calibration source at the edges.
We created related response matrices and ancillary response files
using the tools {\textsc{xisrmfgen}} and {\textsc{xissimarfgen}}.
The {\hxd} spectra were reduced with the tool 
{\textsc{hxdpinxbpi}}. The tuned background 
files\footnote{We used a new version 2.2 updated on 2014 Jun 9 (http://heasarc.gsfc.nasa.gov/docs/suzaku/aehp\_data\_analysis.html).}
were used to model the non X-ray background. 
The cosmic X-ray background was estimated
using the model by \citet{Boldt1987}
in accordance with the {{Suzaku Data Reduction Guide}}.

We used the Xspec software \citep{1996ASPC..101...17A} version 12.8.2
for the spectral analysis.
The cross-normalisation factors between the spectra of different {\suzaku} detectors 
were 
free for {\xis}\,1 and {\xis}\,3,
and fixed to 1.16 for {\hxd} according to the Suzaku Data Reduction Guide. 
Fits were performed in the 0.5--10\,keV energy interval
for the {\xis} spectrum, and from 15\,keV for the {\hxd} spectrum. 
We fit the {\hxd} spectra up to the energy corresponding to 
a 5$\%$ source detection level above the background level which is well above the reported
systematic uncertainty in the {{Suzaku Data Reduction Guide}}.
All {\hxd} spectra but the second one were considered in the 15--55\,keV energy range.
Only the {\hxd} spectrum of the second observation was limited 
by 45\,keV due to a lower signal-to-noise ratio
caused by a shorter exposure time ($\approx$\,22\,ks) 
compared to the other observations. 

We used C-statistics \citep{Cash1979} 
for fitting the data.
However, we also express the goodness of the fit 
with more familiar $\chi^{2}$ values
that were used as a test statistics. Only for this purpose, 
we binned the spectra to contain at least 30 counts per bin.
All the quoted errors correspond to a $90\%$ confidence level
for one interesting parameter.

\begin{figure*}[tb]
 \includegraphics[width=0.49\textwidth]{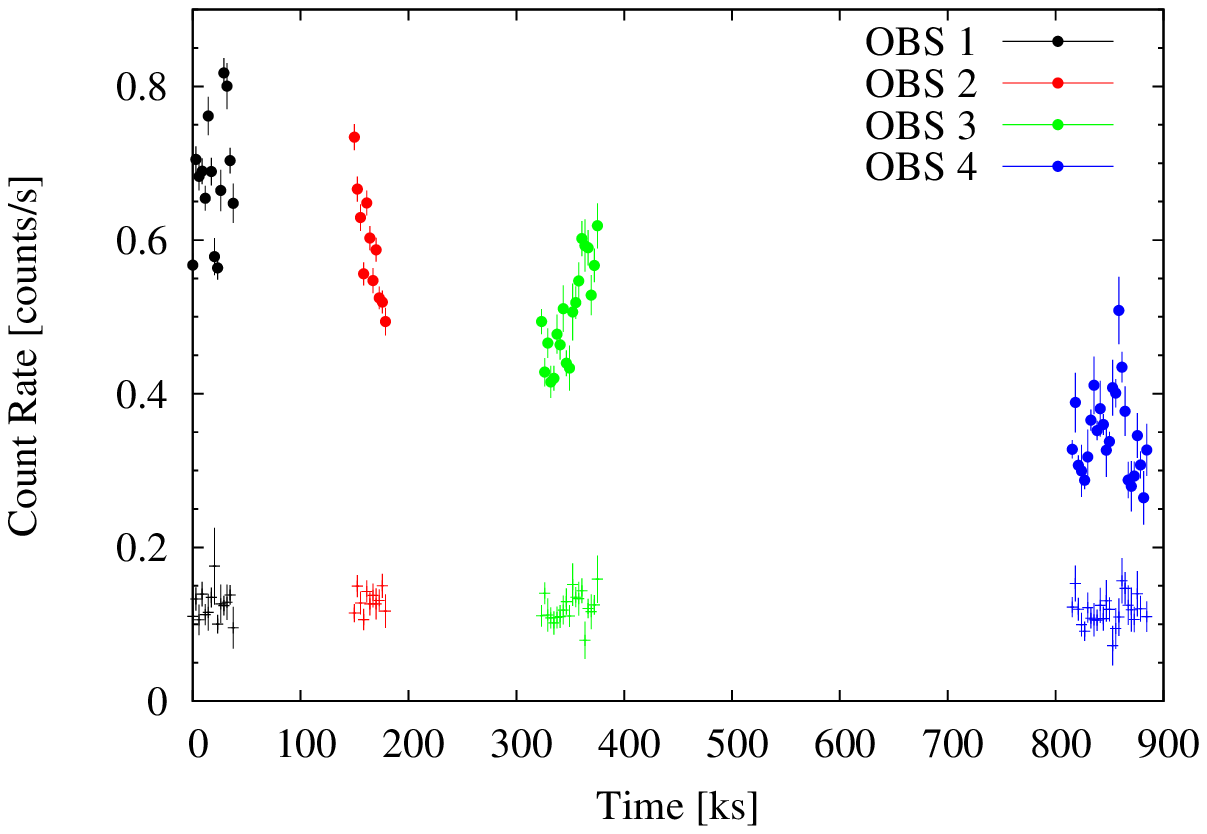}
  \includegraphics[width=0.49\textwidth]{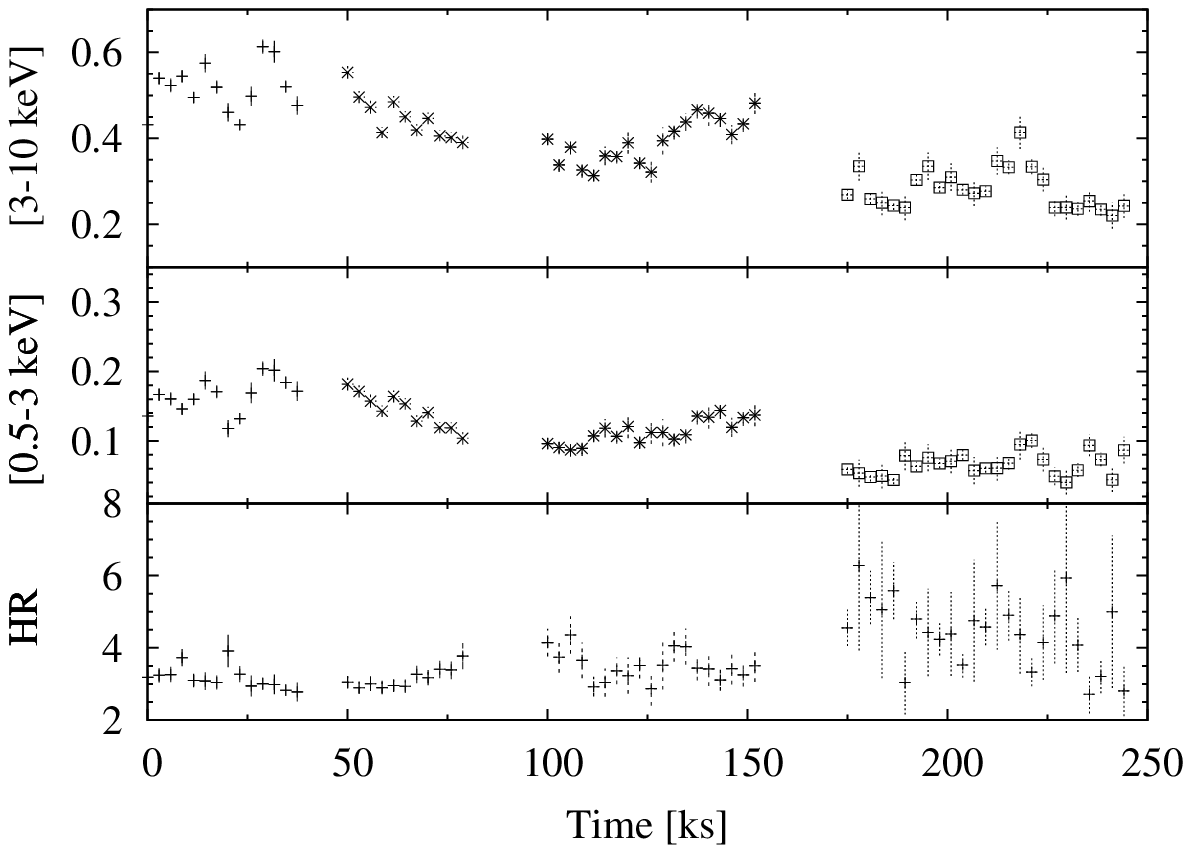}
 \caption{
{\textbf{Left:}}  {\xis}\,0 (upper, with small circles) and {\hxd} (lower)
(all background subtracted) light curves
in the $0.5-10$\,keV, or $15-55$\,keV, energy range,
respectively. 
{\textbf{Right:}} The {\xis}\,0 light curves in two different energy bands
0.5--3\,keV and 3--10\,keV (all background subtracted), and the hardness
ratio between the hard and the soft band. The gaps between the observations are
shrunk for the clarity. The time bin size is 2880\,s.
}
\label{fig_timing}	
\end{figure*}

\section{Results}
\label{results}

%

\subsection{Timing properties}
\label{timing}

We performed a series of four observations separated by
1.5, 2 and 5.5 day intervals to constrain the characteristic time scale of the 
short-term variability of {\source}. 
Figure~\ref{fig_timing} (left panel)
shows the XIS0 and HXD/PIN light curves
from all four observations. While the soft X-ray flux 
decreased by a factor of two between the first and the last observation,
the hard X-ray flux stays more or less constant.

The soft X-ray variability is shown in more details in 
the right panel of Fig.~\ref{fig_timing}
where the light curves in two energy bands, 0.5--3 and 3--10\,keV,
are shown together with the hardness ratio defined as the flux ratio
between these two bands (hard to soft).
The major change in the hardness is measured during the last observation,
when the hardness ratio is significantly larger
than in the previous three observations
suggesting a significant change of the spectrum during this observation.

The flux significantly varied also within the individual observations.
It decreased almost to its half during the second observation,
but the source became brighter again during the third observation.
This variability is accompanied with a small hump in the hardness ratio.
However, a low signal-to-noise ratio limit prevented us from
a detailed investigation of this feature in the light curve,
and we further deal with four time-averaged spectra of the individual observations.

%

\begin{figure*}[tb]
\centering
 \includegraphics[width=0.5\textwidth]{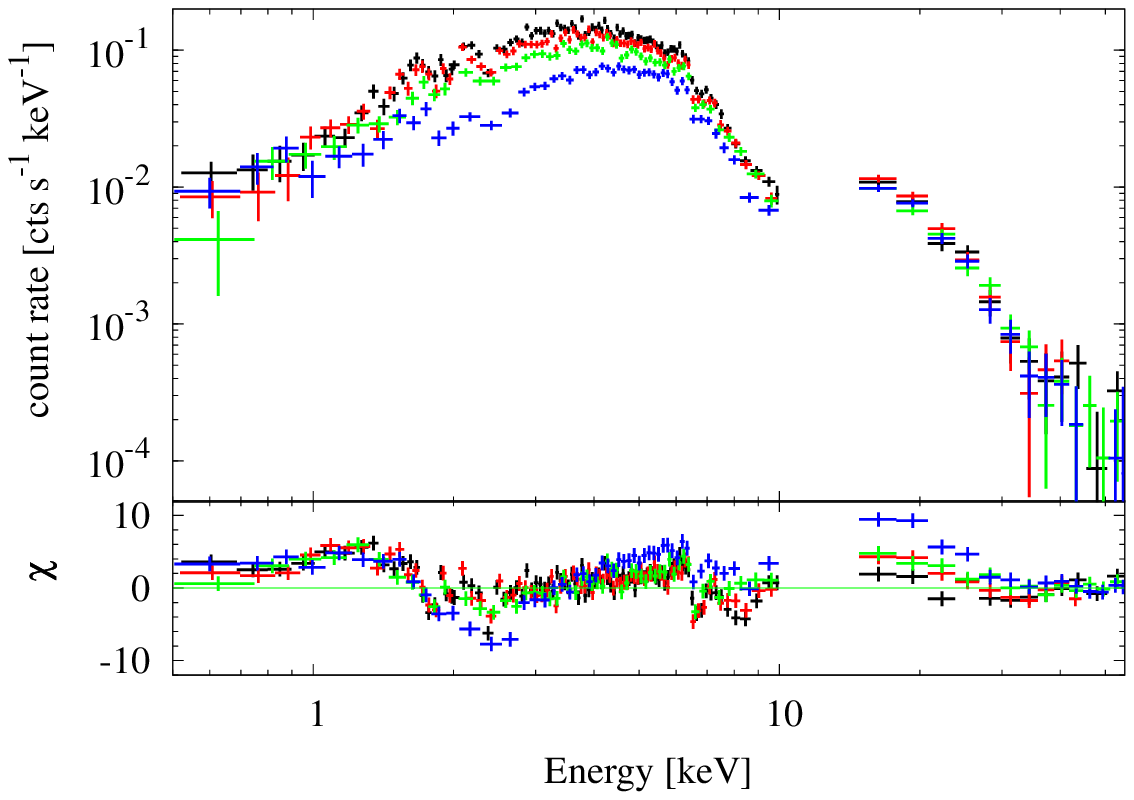}
 \includegraphics[width=0.49\textwidth]{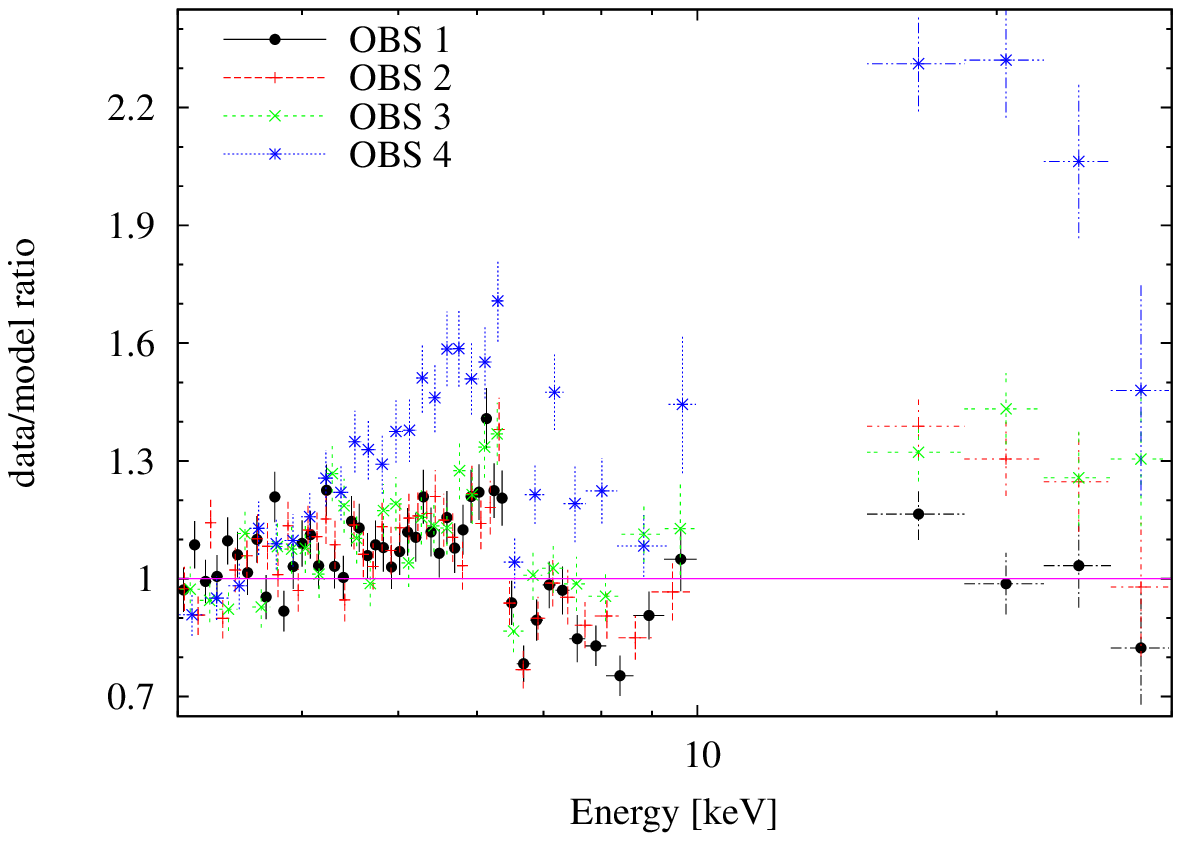}
 \caption{
 {\textbf{Left:}} 
 {\textit{Up:}} X-ray spectra of four {\suzaku} observations (1st black,
 2nd red, 3rd green, 4th blue).
 {\textit{Bottom:}}
 Data residuals from a simple absorbed power-law model with $\Gamma \approx 1.7$.
 {\textbf{Right:}}
 A more detailed look at the iron line and Compton hump
 as residuals from the simple absorbed power-law model.
 Only {\xis}\,0 and {\hxd} spectra are plotted for clarity. Data are re-binned for plotting purposes only.
 }
\label{spectrum}	
\end{figure*}

\subsection{First look at X-ray spectra}

To have a first look at the spectra,
we employed a simple model consisting of a power law 
with its normalisation as the only variable parameter
between the observations. The power law was
affected by the photoelectric absorption by cold material,
for which we used the {\tbabs} model \citep{wilms00}
with Solar abundances.
The column density was a free parameter but linked
between the four observations.

Figure~\ref{spectrum} shows the X-ray spectra of all {\suzaku} observations 
together with the data residuals from this simple model.
The power-law photon index is $\Gamma \approx 1.7$ and 
the absorber's column density is $N_{\rm H} \approx 3.8 \times 10^{22}$\,cm$^{-2}$.
The quality of the fit   
is not acceptable with $C/\nu = 12033/6454 \approx 1.9$
and with clear data residuals from the model
over the entire energy range (see the left bottom panel of Fig.~\ref{spectrum}).


The most evident spectral variability is at 2--6\,keV,
while the spectra do not seem to vary at very high and also very soft ($\lesssim$ 1\,keV) X-ray energies. 
The latter suggests that the spectrum is dominated by two different spectral components 
in the 0.5--1 and 2--6\,keV  energy bands. 
A dominating spectral component at the very low energies can be 
a scattered power-law emission
that is not affected by a variable absorption
of a circumnuclear matter that strongly affects the direct emission,
similarly to what has been found for other similar sources, e.g., ESO 323-G77 \citep{Miniutti2014}.

Other data residuals from the simple power-law model occur at the iron line band
(see the right panel of Fig.~\ref{spectrum}).
The emission line profile itself seems to be asymmetrically broadened
towards its red wing, especially in the lowest-flux observation. 
However, it would be premature to draw
any conclusions about the iron line before finding 
an appropriate description for the continuum.
Therefore, we start with a Gaussian model for the iron line.
Beyond the emission iron line, two absorption features
are visible in the spectra.

The absorption features above the iron line are likely due to 
an ionised gas in the line-of-sight,
probably associated to the warm absorber
that was already revealed by \citet{ricci2010}  
who identified in total three ionised absorbers
with different ionisation based on the {\xmm} spectral analysis.

The simple power-law model fails to describe the hard X-ray emission
(see the right panel of Fig.~\ref{spectrum}).
The spectral curvature might be explained by a reflection hump.
Its presence close to the lower energy limit of the {\hxd} spectrum
suggests that the reflection emission is red-shifted.

Finally, the spectral features around 2\,keV might be affected by calibration
uncertainties and therefore we further omit the data at 1.6-2.4\,keV.

Summarising the results of the first-look analysis,
{\source} is variable mainly at the 2--6\,keV energy band
while there is only little variability in the hard X-rays and also around 1\,keV.

\begin{sidewaystable*} 
\begin{center}
{\small
\caption{Spectral analysis of {\source} - model parameters.}
\label{model}
\begin{tabular}{c|c|cc|cc|cc|cc}
 \rule{0cm}{0.5cm}
  Model Component & Parameter & \multicolumn{2}{c|}{Model A} &  \multicolumn{2}{c|}{Model B}  &  \multicolumn{2}{c|}{Model C}  &  \multicolumn{2}{c}{Model D}	 \\
   \rule{0cm}{0.5cm}
  &  & 1st obs & 4th obs	 & 1st obs & 4th obs	 & 1st obs & 4th obs	 & 1st obs & 4th obs	 \\
\hline \hline
\rule[-0.7em]{0pt}{2em} 
power law & $\Gamma$  & \multicolumn{2}{c|}{$1.84^{+0.03}_{-0.02}$} & \multicolumn{2}{c|}{$1.80 \pm 0.02$} & \multicolumn{2}{c|}{$1.96^{+0.04}_{-0.01}$} & \multicolumn{2}{c}{$2.26\pm0.04$} \\
\rule[-0.7em]{0pt}{2em} 
... & $K_{\rm direct} [10^{-3}]$ & $11.5^{+0.4}_{-1.1}$ & $10.6^{+1.0}_{-0.4}$ &  $11.0^{+0.3}_{-0.5}$  & $9.9^{+0.5}_{-0.4}$  & $13.7^{+1.0}_{-0.3}$ & $11.0^{+0.8}_{-0.3}$ & $20.7^{+1.4}_{-2.0}$ & $10.1^{+0.7}_{-0.9}$ \\
\rule[-0.7em]{0pt}{2em} 
... & $K_{\rm scattered} [10^{-3}]$ & $0.19^{+0.01}_{-0.02}$ & $0.17 \pm 0.01$ & $0.18^{+0.02}_{-0.01}$ & $0.17 \pm 0.01$ &  $0.12^{+0.02}_{-0.03}$ & $0.13^{+0.02}_{-0.01}$ & $0.13 \pm 0.02$ & $0.14^{+0.01}_{-0.02}$ \\
\hline
\rule[-0.7em]{0pt}{2em} 
local cold absorber & $N_{\rm H} [10^{22}$\,cm$^{-2}]$ & $3.9\pm0.1$ & $7.6\pm0.4$ & $3.7^{+0.2}_{-0.1}$ & $7.1^{+0.4}_{-0.3}$ & \multicolumn{2}{c|}{-}  & \multicolumn{2}{c}{-} \\
\hline
\rule[-0.7em]{0pt}{2em} 
low-ionised absorber  & $N_{\rm H} [10^{22}$\,cm$^{-2}]$ & \multicolumn{2}{c|}{-} & \multicolumn{2}{c|}{-} & $2.7^{+0.5}_{-1.0}$ & $5.7 \pm 0.4$ & $2.9^{+0.7}_{-0.4}$ & $6.7^{+1.1}_{-1.4}$ \\
\rule[-0.7em]{0pt}{2em} 
... & $\log\xi$ & \multicolumn{2}{c|}{-} & \multicolumn{2}{c|}{-} & \multicolumn{2}{c|}{$1.1^{+0.1}_{-0.2}$} & \multicolumn{2}{c}{$1.2 \pm 0.1$} \\
 \hline
\rule[-0.7em]{0pt}{2em} 
mid-ionised absorber 2 & $N_{\rm H} [10^{22}$\,cm$^{-2}]$ & \multicolumn{2}{c|}{-} & \multicolumn{2}{c|}{-} & $7.3^{+1.3}_{-0.5} $ & $8.2^{+0.7}_{-4.2}$ & $7.7 \pm 0.4$ & $6.5^{+1.6}_{-1.5}$ \\
\rule[-0.7em]{0pt}{2em} 
 ... & $\log\xi$ & \multicolumn{2}{c|}{-} & \multicolumn{2}{c|}{-} & \multicolumn{2}{c|}{$1.9 \pm 0.1$} & \multicolumn{2}{c}{$1.6 \pm 0.1$} \\
\hline
\rule[-0.7em]{0pt}{2em} 
high-ionised absorber & $N_{\rm H} [10^{22}$\,cm$^{-2}]$ & $26^{+6}_{-4}$ & $114 \pm 17$ & $18^{+7}_{-5}$ & $98^{+7}_{-5}$ & \multicolumn{2}{c|}{$23^{+15}_{-10}$} & \multicolumn{2}{c}{$28^{+14}_{-7}$} \\
\rule[-0.7em]{0pt}{2em} 
 ... & $\log\xi$ & \multicolumn{2}{c|}{$3.3 \pm 0.1$} & \multicolumn{2}{c|}{$3.2^{+0.1}_{-0.2}$} & \multicolumn{2}{c|}{$3.6^{+0.2}_{-0.1}$} & \multicolumn{2}{c}{$3.6 \pm 0.2$} \\
 \rule[-0.7em]{0pt}{2em} 
 ... & $v_{\rm outflow}$\,[km\,s$^{-1}$] & \multicolumn{2}{c|}{$3300\pm600$} & \multicolumn{2}{c|}{$4200^{+1300}_{-1400}$} & \multicolumn{2}{c|}{$0 \pm 1500$} & \multicolumn{2}{c}{$1400^{+400}_{-700}$} \\
\hline
\rule[-0.7em]{0pt}{2em} 
very-high ionised absorber & $N_{\rm H} [10^{22}$\,cm$^{-2}]$ & \multicolumn{2}{c|}{-} & {$23^{+27}_{-10}$} & {$8^{+13}_{-5}$} & \multicolumn{2}{c|}{$23^{+7}_{-3}$} & \multicolumn{2}{c}{-} \\
\rule[-0.7em]{0pt}{2em} 
 ... & $\log \xi$ & \multicolumn{2}{c|}{-} & $4.2 \pm 0.2$ & $3.6^{+0.5}_{-0.3}$ & \multicolumn{2}{c|}{$4.33^{+0.09}_{-0.15}$} & \multicolumn{2}{c}{-} \\
 \rule[-0.7em]{0pt}{2em} 
 ... & $v_{\rm outflow}$\,[km\,s$^{-1}$] & \multicolumn{2}{c|}{-} & \multicolumn{2}{c|}{$300^{+2300}_{-400}$}  & \multicolumn{2}{c|}{$2700^{+800}_{-1600}$} & \multicolumn{2}{c}{-} \\
\hline
\rule[-0.7em]{0pt}{2em} 
Gaussian line & $E [$keV$]$ & \multicolumn{2}{c|}{$6.37\pm0.02$} & \multicolumn{2}{c|}{$6.38^{+0.03}_{-0.02}$} & \multicolumn{2}{c|}{-} & \multicolumn{2}{c}{-} \\
\rule[-0.7em]{0pt}{2em} 
 ... & $\sigma [$keV$]$ & \multicolumn{2}{c|}{$0.03^{+0.04}_{-0.03}$} & \multicolumn{2}{c|}{$0.04^{+0.06}_{-0.04}$} & \multicolumn{2}{c|}{-}& \multicolumn{2}{c}{-} \\
 \rule[-0.7em]{0pt}{2em} 
 ... & normalisation $[10^{-5}]$ & \multicolumn{2}{c|}{$1.2\pm0.2$} & \multicolumn{2}{c|}{$1.2^{+0.5}_{-0.3}$} & \multicolumn{2}{c|}{-}& \multicolumn{2}{c}{-} \\
  \rule[-0.7em]{0pt}{2em} 
 ... &  equivalent width [eV]& $35^{+8}_{-5}$ & $56^{+11}_{-8}$ & $36^{+16}_{-9}$ & $58^{+8}_{-14}$ & \multicolumn{2}{c|}{-}& \multicolumn{2}{c}{-} \\
 \hline
 \rule[-0.7em]{0pt}{2em} 
relativistic reflection  & spin $a$/$M$ & \multicolumn{2}{c|}{-}& \multicolumn{2}{c|}{-}& \multicolumn{2}{c|}{-}& \multicolumn{2}{c}{$0.80^{+0.11}_{-0.15}$} \\
 \rule[-0.7em]{0pt}{2em} 
 ... & inclination [deg] & \multicolumn{2}{c|}{-}& \multicolumn{2}{c|}{-}& \multicolumn{2}{c|}{-}& \multicolumn{2}{c}{$20^{+6}_{-20}$} \\
  \rule[-0.7em]{0pt}{2em} 
... & radial emissivity & \multicolumn{2}{c|}{-}& \multicolumn{2}{c|}{-}& \multicolumn{2}{c|}{-}& \multicolumn{2}{c}{$3.5 \pm 0.3$} \\ 
 \rule[-0.7em]{0pt}{2em} 
 ... & normalisation $[10^{-3}]$& \multicolumn{2}{c|}{-}& \multicolumn{2}{c|}{-}& \multicolumn{2}{c|}{-}&  $54^{+20}_{-32}$ & $71^{+25}_{-30}$ \\
 \hline
 \rule[-0.7em]{0pt}{2em}
distant reflection  & iron abundance [Solar] & \multicolumn{2}{c|}{-}& \multicolumn{2}{c|}{-}& \multicolumn{2}{c|}{$<0.35$} & \multicolumn{2}{c}{$0.89^{+0.18}_{-0.14}$} \\
 \rule[-0.7em]{0pt}{2em} 
 ... & normalisation $[10^{-3}]$& \multicolumn{2}{c|}{-}& \multicolumn{2}{c|}{-}& \multicolumn{2}{c|}{$6.3^{+1.0}_{-0.6}$} & \multicolumn{2}{c}{$9.4^{+2.5}_{-1.4}$} \\
 \hline
\rule[-0.7em]{0pt}{2em} 
fit goodness  & $C/\nu$ & \multicolumn{2}{c|}{$6399/5809$} & \multicolumn{2}{c|}{$6297/5801$} &  \multicolumn{2}{c|}{$6094/5810$} & \multicolumn{2}{c}{$5988/5806$} \\
test statistics & $\chi^2/\nu$ & \multicolumn{2}{c|}{$6410/5809$} & \multicolumn{2}{c|}{$6294/5801$} & \multicolumn{2}{c|}{$6097/5810$} & \multicolumn{2}{c}{$5967/5806$} \\
\end{tabular}
\tablefoot{The values of the model parameters which were allowed to vary
between the four observations 
are reported for the 1st (highest-flux) and 4th (lowest-flux) observations only for brevity.
A single value for all observations is reported
when the parameters were free in fitting but fixed to the same value
or did not significantly change
between the observations.
All the errors correspond to the 90\% confidence level for one interesting parameter.
All normalisations are expressed in units of photons\,keV$^{-1}$\,cm$^{-2}$\,s$^{-1}$ at 1\,keV.}
}
\end{center} 
\end{sidewaystable*} 


\begin{figure}[tb]
\includegraphics[width=0.49\textwidth]{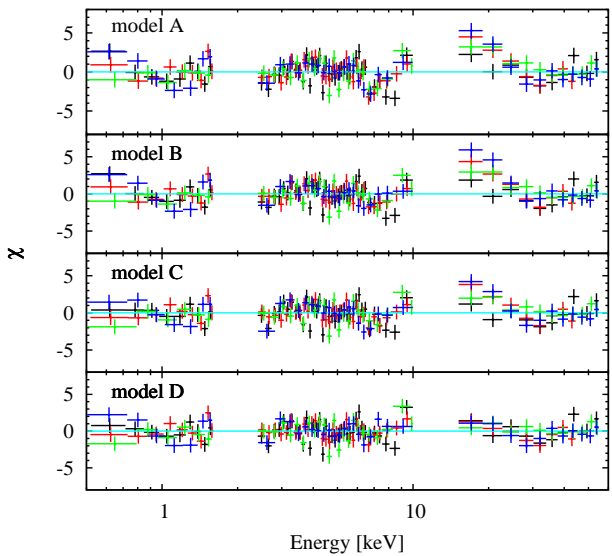}
\caption{Data residuals from the different models for all observations
 (1st black, 2nd red, 3rd green, 4th blue).
{\textbf{From top to bottom:}} Residuals from the model A--D.
}
\label{pldelc}
\end{figure}

\subsection{Modelling the complex absorption}

Next, we employed a model with
two power-law components with the photon index linked between them -- 
one affected by the local absorber and one not.
The local absorber was first modelled by
a cold absorber using {\ztbabs} model
with the column density free to vary between the observations.
We further used two Gaussian lines
to model the absorption features in the iron-line energy band,
with free energies and the redshift fixed to the cosmological redshift of the galaxy.
Furthermore, we used a Gaussian line profile to account for the emission iron K$\alpha$ line.
The global model was then absorbed
by the interstellar absorption in our Galaxy with the column density
$N_{\rm H} = 6.9 \times 10^{20}$ cm$^{-2}$ \citep{Kalberla2005} using the {\tbabs} model.
Solar abundances were assumed in all absorption models.

We obtained a statistically better fit but not yet acceptable
with $C/\nu = 7730/5798 \approx 1.3$.
The column density of the variable absorber was found to increase
from $N_{\rm H} = 2.95^{+0.08}_{-0.07} \times 10^{22}$ cm$^{-2}$ during the 1st observation
to $N_{\rm H} = 6.8 \pm 0.3 \times 10^{22}$ cm$^{-2}$ during the 4th observation.
For the iron K$\alpha$ emission line, we obtained
$E = 6.32 \pm 0.03$\,keV, $\sigma = 0.14 \pm 0.03$\,keV and equivalent width
ranging from $EW = 120^{+50}_{-30}$\,eV (1st observation) to $EW = 320 \pm 40$\,eV (4th observation).
The line was found to be broadened and red-shifted from its intrinsic 6.4\,keV energy.
The FWHM widths of the absorption lines were found even larger. Their values were pegged 
at the upper limits, which we set to $\sigma = 0.3$\,keV. The energy of the first absorption line 
was found to be $E = 6.75 \pm 0.1$\,keV for the first three observations
and $E = 6.6 \pm 0.1$\,keV for the 4th observation.
The energy of the second absorption line 
was found to be $E = 8.2 \pm 0.2$\,keV for all observations,
consistent within the uncertainties.

The power-law photon index in this simple model was $\Gamma \approx 1.5$.
Unusually flat spectral slopes have been often found to be due to a partially-covering 
ionised absorber \citep[see e.g.][]{Piconcelli2004, Mathur2009, 2012A&A...545A.148S}.
As a next step, we therefore employed a partially-covering ionised absorption
to model the absorption features in the spectra.
We replaced the absorption Gaussian lines
by the {\zxipcf} model \citep{Reeves2008}
using the {\xstar} photo-ionisation code \citep{xstar}.
We first start with one component and refer to this model as ``model A''.
The best-fit values are presented in Table~\ref{model}
and the data residuals from the model are shown in the upper panel of Fig.~\ref{pldelc}. 
The photon index of the power law $\Gamma \approx 1.8$ is consistent
with the characteristic slope of Seyfert~1 galaxies \citep[see e.g.][]{2009A&A...495..421B, Malizia2014}.
We found high ionisation of the absorber, $\log \xi \gtrsim 3$, 
consistently with \citet{ricci2010},
but with the column density $N_{\rm H} \approx 10^{23}-10^{24}$\,cm$^{-2}$,
roughly an order of magnitude larger.
However, the global fit is not satisfactory with $C/\nu \approx 1.1$,
and, in particular, residuals around $\approx 7$\,keV indicate 
the need for another high-ionised absorption component.

The addition of another warm absorber component 
significantly improved the fit (from $C/\nu = 6399/5809$ to $6297/5801$). 
The statistical F-test 
using the $\chi^2$ test statistics gives the probability
$\approx 3\times10^{-19}$ that the improvement is just coincidental.
We refer to this model as ``model B''
and the best-fit values are summarised in the Table~\ref{model}.
The value of the ionisation of the second warm absorber is higher, $\log\xi \approx 4$,
and it accounts for the residuals above 7\,keV (see the second panel of Fig.~\ref{pldelc}).
We allowed also the covering factors to vary,
and we tried several tests with different initial values,
but we found no statistical improvement compared to fits
when these parameters were fixed to 1.
The ionisation parameter was also left free
but no significant change was detected across the observations. 
Only column densities of both cold and warm absorber
were required to vary significantly.

\begin{figure}[tb]
\includegraphics[width=0.49\textwidth]{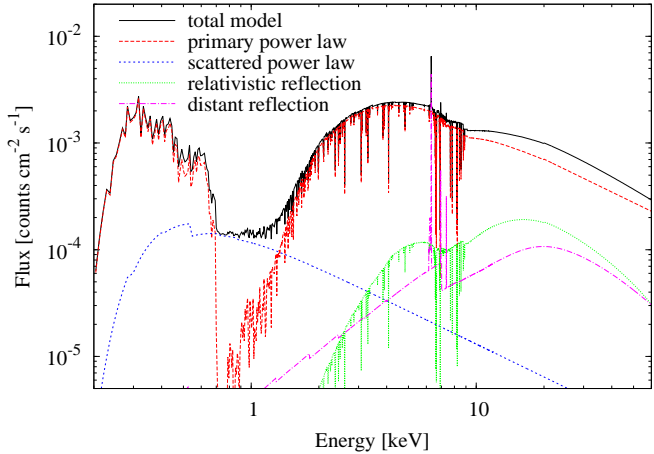}
\caption{The final model (model D) with its components.
}
\label{plot_model}	
\end{figure}

\begin{figure}[tb]
\includegraphics[width=0.49\textwidth]{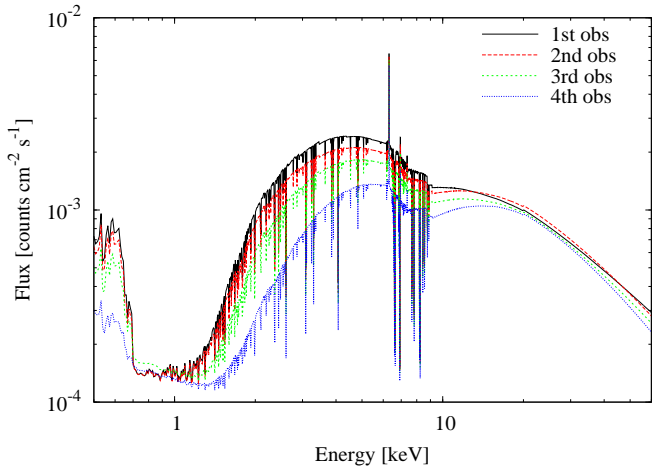}
\caption{The final model (model D) in the individual observations (1st black, 2nd red, 3rd green, 4th blue).
}
\label{plot_model_observations}	
\end{figure}

As a next step, we allowed the cold absorber to be ionised and partially covering.
In XSPEC, we replaced the {\ztbabs} model by another {\zxipcf} component.
We first allowed also the redshift of this absorber to vary.
However, its value reached 0.1, 
which would correspond to an inflow of $\gtrsim 25000$ km\,s$^{-1}$. 
This is a quite extreme value that has not been reported in any similar source.
Because the redshift in this case
is not obtained from the frequency shift of a clearly defined absorption line, we considered this measurement
as not indicative and we rather fix the redshift to the cosmological value of the galaxy.
The obtained fit is significantly better than the previous one with
$C/\nu = 6228/5804$, i.e., $\Delta C/\nu = 69$.
However, we realised that instead of having
two highly-ionised and one low-ionised component, 
the fitting procedure converged to
a low-ionised, mid-ionised and one highly-ionised component. 
Contrary to the previous model, the column densities of the high-ionised warm absorber
were not required to vary between observations.
Also, we found that the ionisation
was not required to vary, 
and covering factors were consistent with being 1 for all observations.
As a check of this result, we linked
the column densities, and checked whether variable covering factors 
and ionisations may describe the data as well. However, we obtained
statistically worse fit with $C/\nu = 6259/5795$.

We checked the residuals and realised that the most
significant improvement compared to model B 
appeared at energies $\lesssim 1$\,keV,
while the residuals at the iron line energy band got worse.
We therefore added another absorber to account again for high ionisation.
We found a new statistically better fit with four absorbers
($C/\nu = 6138/5797$, $\Delta C/\nu = 90$
from the previous model).
However, we did not obtain any significant improvement
in modelling the residuals at energies above 10\,keV.

\subsection{Modelling the reflection}

The models presented so far were not able to fully describe
the data residuals in the iron line band and around 15\,keV.
The presence of iron emission line is required in all models,
which indicates a significant contribution of X-ray reflection to the observed spectrum.
To consistently treat the X-ray reflection,
we replaced the Gaussian line by the reflection model {\pexmon} \citep{nandra2007}
that combines the iron line and the reflection continuum using the {\pexrav} model \citep{pexrav}.
We used the {\pexmon} model only for reflection, i.e., with the reflection fraction parameter $R$ fixed to $-1$.\footnote{Only 
to estimate the reflection strength we also used the {\pexmon} model 
for primary power-law emission. We fixed the power-law normalisation
to zero and let the parameter $R$ to vary.
We got the reflection strength $R \approx 0.5$.}
The inclination was fixed to 45 degrees.
We refer to this model as ``model C''
and the best-fit values are presented in the Table~\ref{model}.
The statistical improvement compared to the previous model
is $\Delta C/\nu = 43$.
However, the addition of the distant reflection component 
still did not improve satisfactorily
the most prominent residuals at the {\hxd} energy band
(see the third panel of Fig.~\ref{pldelc}).

As a next step, we employed reflection
from the innermost region of the accretion disc
where the radiation is significantly red-shifted due to effects of the general relativity. 
We therefore added a second reflection component
to the model that is convolved with a relativistic disc kernel \citep{ky}.
Free parameters of the relativistic model
were the black hole spin, inclination angle and radial-emissivity index. 
The inner radius
was set to the marginally stable orbit depending on the spin value.
The outer radius was set to 400\,$r_{\rm g}$, where $r_{\rm g}=\frac{GM}{c^2}$
is the gravitational radius. The iron abundances were linked to the abundances
of the distant reflection model. We still kept the inclination of the distant reflection
fixed to 45 degrees, because the global fit is not very sensitive to this parameter.

We further assumed that the relativistic reflection is affected by all absorbers
while the neutral reflection coming possibly from the distant polar scattering region
is unaffected by the warm absorber.
We obtained a significantly improved fit ($C/\nu = 5988/5806$, $\Delta C/\nu = 106$ from the previous model). 
The best-fit values are presented in the Table~\ref{model} as model ``D'' parameters.
We considered only three absorbers in this model,
because we realised during the fitting procedure
that the highest ionised absorber becomes less important with the new model
(the difference between the model with three and four absorbers was only
$\Delta C \approx 10$).
Some of the absorption features at 7--10\,keV were likely attributed
to absorption edges of the reflection models.

The final model is statistically well acceptable with no significant residuals
across the entire {\suzaku} bandpass (see the bottom panel of  Fig.~\ref{pldelc}).
The model with its components is shown in Fig.~\ref{plot_model}.
Figure~\ref{plot_model_observations} shows the comparison of the final model in all four observations.
The low-ionised absorber is the most responsible for the curvature of the spectrum
at energies 1--10\,keV. The scattered power-law emission dominates at 1\,keV
over the absorbed intrinsic flux and implies that the flux is not strongly variable there. 
The hard energy band is shaped by reflection. Relativistic smearing is responsible
for shifting the Compton hump to lower energies.

We note a minor discrepancy between the data residuals and model predictions
in the lower edge of the {\xis} spectrum,
compare Fig.~\ref{spectrum} and Fig.~\ref{plot_model_observations} at 0.5--0.7\,keV.
This can be, however, an instrumental effect due to the redistribution
of the more energetic photons to lower energies that appears
at the low edge of the spectrum \citep{Koyama2007}.



\section{Discussion}
\label{discussion}

\begin{table*}
{\small
\caption{Observed and intrinsic flux and luminosity measurements of {\source}.}
\label{table_luminosities} 
\centering
\begin{tabular}{ccccccc}
\rule{0cm}{0.5cm}

observation & F$_{2-10\,{\rm keV, observed}}$ & F$_{2-10\,{\rm keV, intrinsic}}$ \
& L$_{2-10\,{\rm keV, observed}}$ & L$_{15-50\,{\rm keV, observed}}$ &  L$_{\rm bolometric}$ \
&  L$_{0.0136-13.6\,{\rm keV, intrinsic}}$\\
\rule[-0.7em]{0pt}{2em} 
number & [$10^{-11}$\,erg\,cm$^{-2}$\,s$^{-1}$] & [$10^{-11}$\,erg\,cm$^{-2}$\,s$^{-1}$] \
&[$10^{43}$\,erg\,s$^{-1}$] &[$10^{43}$\,erg\,s$^{-1}$] &[$10^{44}$\,erg\,s$^{-1}$] &[$10^{44}$\,erg\,s$^{-1}$] \\
\hline \hline
\rule[-0.7em]{0pt}{2em} 
1 & $2.3 \pm 0.1$ & $4.1 \pm 0.4$ & $1.8 \pm 0.2$ &  $1.9 \pm 0.2$ & $3.7 \pm 0.4$ & $1.5 \pm 0.2$ \\
\rule[-0.7em]{0pt}{2em} 
2 & $2.1 \pm 0.1$ &  $3.6 \pm 0.4$ & $1.6 \pm 0.2$ & $1.9 \pm 0.2$ & $3.2 \pm 0.4$ & $1.3 \pm 0.2$ \\
\rule[-0.7em]{0pt}{2em} 
3 & $1.8 \pm 0.1$ &  $3.1 \pm 0.3$ & $1.4 \pm 0.1$ & $1.8 \pm 0.2$ & $2.7 \pm 0.3$ & $1.1 \pm 0.1$ \\
\rule[-0.7em]{0pt}{2em} 
4 & $1.3 \pm 0.1$ &  $2.5 \pm 0.3$ & $1.1 \pm 0.1$ & $1.7 \pm 0.2$ & $2.0 \pm 0.3$ & $0.8 \pm 0.1$ \\
\end{tabular}
}
\end{table*}

\begin{figure}[tb]
\includegraphics[width=0.49\textwidth]{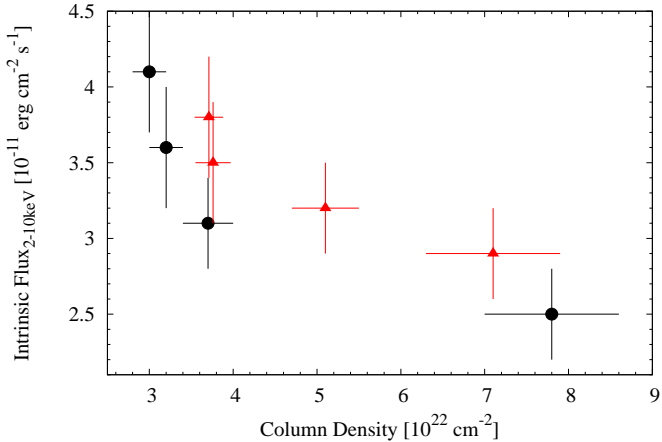}
\caption{Relation between the intrinsic 2--10\,keV flux
and the column density of the cold (least ionised) absorber
for the model D (black, circles) and model B (red, triangles).
}
\label{rel_intflux_abs}	
\end{figure}

\begin{figure}[tb]
\includegraphics[width=0.49\textwidth]{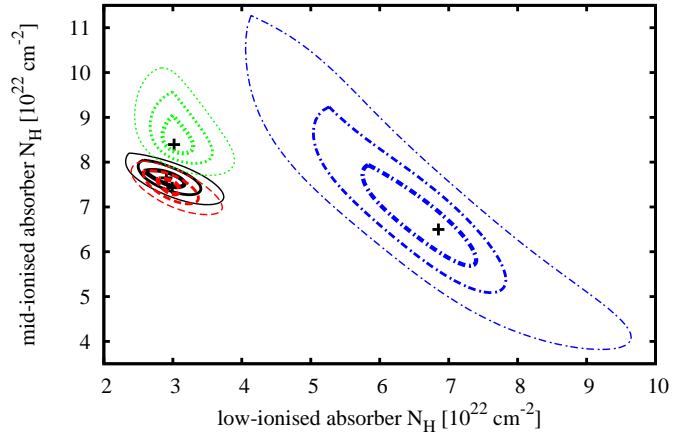}
\caption{Contour plots between the column densities of the variable 
 low- and mid-ionised absorbers. The colours correspond to different observations
 (1st black, 2nd red, 3rd green, 4th blue). The contours correspond to 1$\sigma$,
 2$\sigma$, and 3$\sigma$ levels. The best-fit values are marked by a small cross at the graph.
}
\label{contnh_modelD}	
\end{figure}

\begin{figure*}[tb]
 \includegraphics[width=0.48\textwidth]{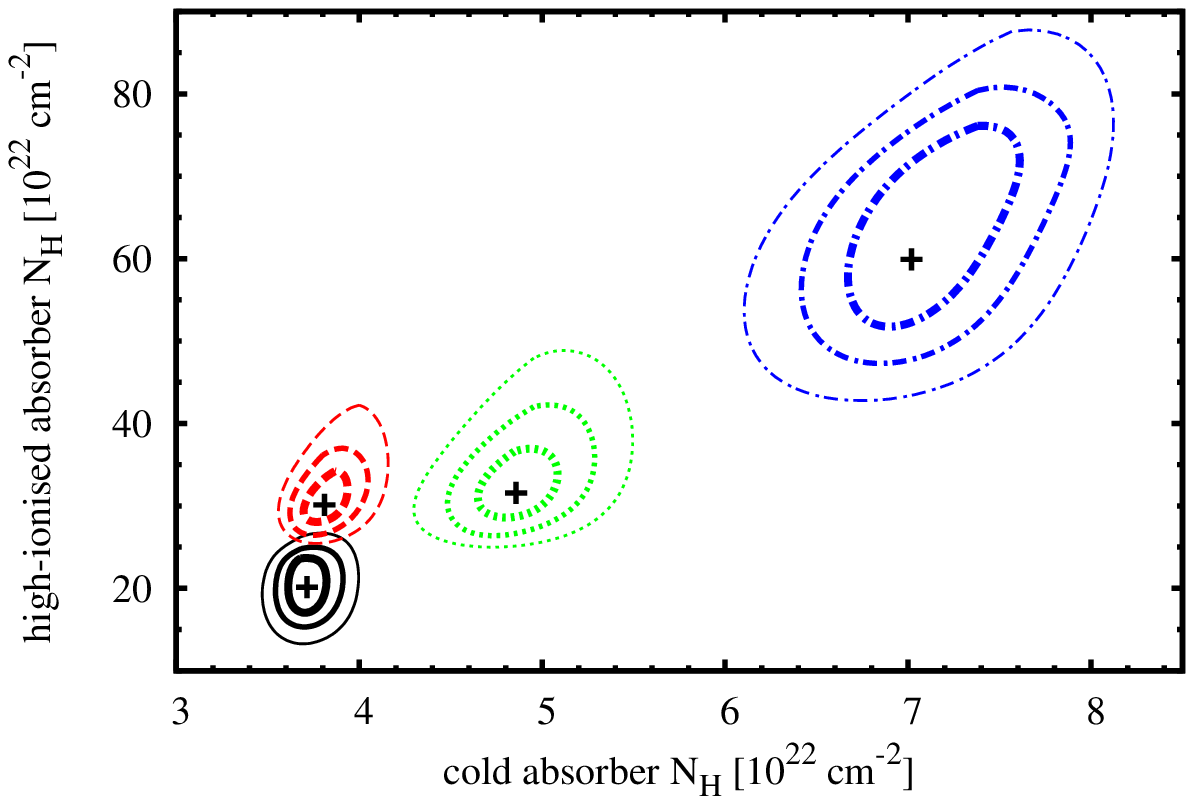}
 \includegraphics[width=0.48\textwidth]{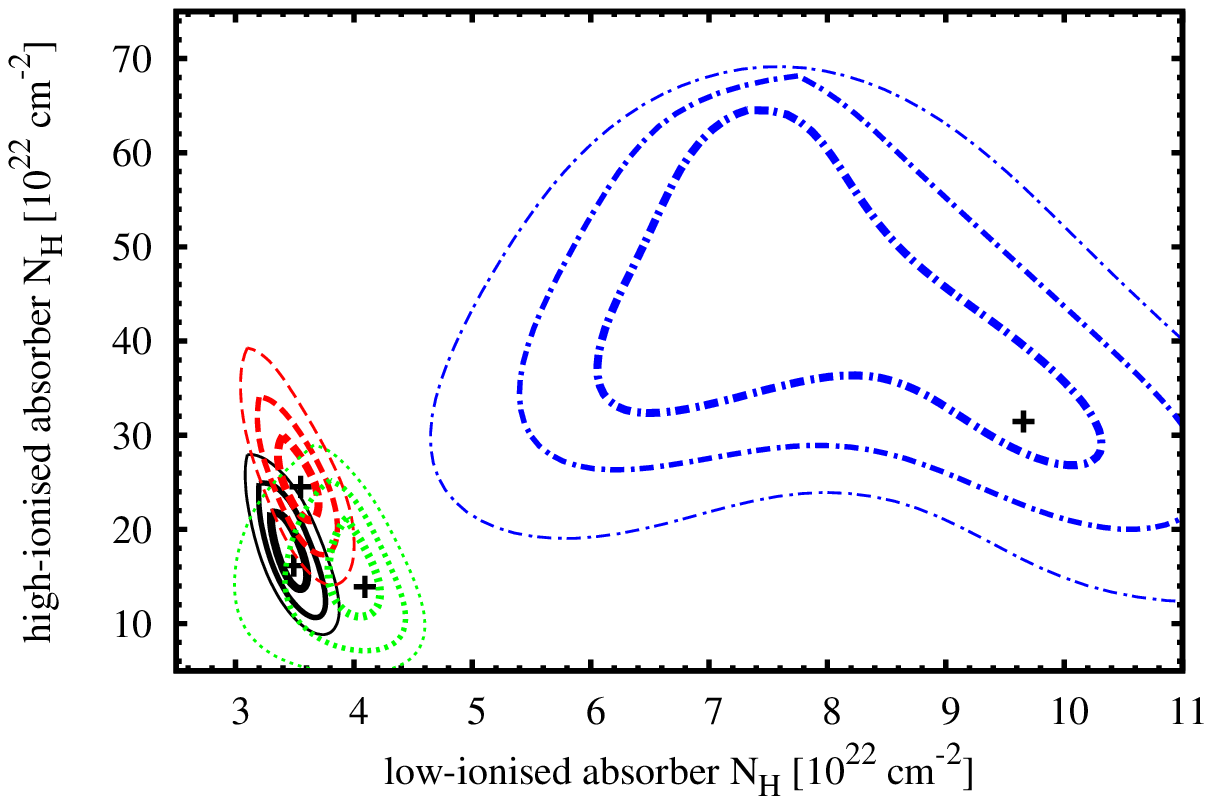}
 \caption{Contour plots between the column densities of the variable 
 cold and warm absorbers. The colours correspond to different observations
 (1st black solid, 2nd red dashed, 3rd green dotted, 4th blue dash-dotted). The contours correspond to 1$\sigma$,
 2$\sigma$, and 3$\sigma$ levels. {\textbf{Left:}} Results obtained with the model B. 
 {\textbf{Right:}} Results obtained with the model D with highly-ionised warm absorber allowed to vary.
 }
\label{contnh}	
\end{figure*}




\subsection{Variable spectral components}
\label{variable}

Our monitoring observational programme revealed
that the short-term variability of {\source} 
is partly caused 
by the variable column density of the absorbing gas,
and partly 
by the changes in
the intrinsic luminosity that also significantly 
varies in the soft X-rays between the observations. 
Table~\ref{table_luminosities} shows our measured values
of the observed X-ray flux in the 2--10\,keV range,
the intrinsic 2--10\,keV fluxes constrained from
our final model, the observed luminosity determined
using the cosmological distance\footnote{using the
cosmological parameters $H_{0} = 70$\,km\,s$^{-1}$\,Mpc$^{-1}$,
$q_0=0$, and $\lambda_0=0.73$},
and the estimation of the bolometric and ionising luminosity.
The bolometric luminosity was constrained from the X-ray luminosity 
using a relation by \citet{Marconi2004}.

Because the intrinsic X-ray luminosity decreased almost by a factor of two
according to the final model, we get different values for the bolometric luminosity.
We note, however, that these are only estimates based on measured X-ray luminosities,
and the bolometric luminosity itself does not need to vary.
The estimated values were $L_{\rm bol}\approx 3.7 \times 10^{44}$\,erg\,s$^{-1}$
for the first observation, and $L_{\rm bol}\approx 2 \times 10^{44}$\,erg\,s$^{-1}$
for the last observation.
This translates to Eddington ratios
$\lambda_{\rm Edd} \approx 0.03$, and $\lambda_{\rm Edd} \approx 0.02$, respectively.
We used relation for the Eddington luminosity as
$L_{\rm Edd} = 1.26 \times 10^{38}\left(M/M_{\odot}\right)$\,erg\,s$^{-1}$
and the value of the mass $M \approx 10^8$\,M$_{\odot}$ \citep{Bennert2006}.

To constrain the ionising luminosity, we used the relation by \citet{Vasudevan2010}
to estimate the soft luminosity $L_{\rm 13.6-100\,eV} = \kappa_{\rm  13.6-100\,eV} L_{\rm bol}$,
where $\kappa_{\rm  13.6-100\,eV}$ goes from $0.21$ to $0.59$
for accretion rates from $0.01$ to $0.61$.
By the linear interpolation, we constrained the factors for each observation
depending on the estimated accretion rate.
We determined the X-ray luminosity $L_{0.1-13.6\,{\rm keV}}$
by an extrapolation of the measured X-ray luminosity in 0.5--10\,keV.
To get the ionising luminosity, we summed the soft and the X-ray luminosity.
The results are shown in the last column of Table~\ref{table_luminosities}.
Figure~\ref{rel_intflux_abs} shows the relation between the column density
of the least-ionised absorber and the intrinsic 2--10\,keV flux. 
This relation depends on the employed model, and therefore
we show this relation for models B and D.
Both results strongly indicate that
the column density of the absorber increases
with decreasing intrinsic flux.

Surprisingly, we have not detected any significant variability
in the ionisation of the absorbers that would reflect
the decrease of the illuminating flux.
This would suggest that the absorber is not in the photo-ionisation equilibrium.
Such behaviour has been reported also for some other sources, such as Mrk~335 \citep{Longinotti2013},
or NGC~3783 \citep{Krongold2005}.
We note, however, that in our case, the models of the low- and mid-ionised absorbers are based
on the continuum spectral shape and not on the individual absorption lines
that would define the ionisation parameter more precisely.
Therefore, no measured variability of the ionisation parameters
may be simply due to coarse resolution of the CCD detectors.

All the employed models suggest that  
the most variable spectral component
is the least-ionised absorber. 
Figure~\ref{contnh_modelD} shows a contour plot between the column densities
of the low- and mid-ionised absorbers in the final model~D.
All the free parameters of the final model were allowed to vary during the contour calculations.
The column density of the low-ionised absorber
changed by a factor of two
between the first three and the last observation,
while the column density of the mid-ionised absorber 
may stay unchanged within 3$\sigma$ levels.
 
As the values of column densities are similar and the ionisation parameters
are not constrained from any discrete features, we tested a scenario,
in which the low- and mid-ionised absorbers are replaced by one variable absorber. 
We considered two cases: (1) the column density and the ionisation are variable between the observations,
and (2) the column density and the covering factor are variable.
In both cases, the resulting fit is worse
by $\Delta{C} \approx 30$ compared to the final model. 
The F-test using $\chi^2$ statistics gives the probability
$p \approx 3\times10^{-6}$ that the statistical improvement with
two absorbers is just coincidental.
Higher-quality data would be needed to more precisely constrain
the structure of the absorbing gas.

Further, we investigated possible variability of the highest ionised absorber.
Figure~\ref{contnh} shows contour plots
between the column density of the cold and highly ionised absorber.
The left panel shows results obtained by the model B.
The right panel shows results obtained by a modified version of the model D,
in which we allowed the warm absorber with high ionisation to vary 
instead of the mid-ionised absorber.
While the warm-absorber column density is required to change in the model B,
the model D shows that this parameter may stay unchanged during all observations.

\subsection{Constraints of the variable absorber's location}
\label{theory}

We found that the spectral change between the last observation
from the previous three observations
is mainly due to the increase of the column density of the cold or least ionised absorber.
Based on the measured time scale of the spectral variability,
we may estimate the location of this variable absorber. 
Let us assume that the main velocity of the absorbing cloud 
is the orbital velocity around the centre that can be approximated
as the Keplerian velocity at the distance $R$:
\begin{equation}
 v_{\rm K} = \sqrt{\frac{GM_{\rm BH}}{R}},
\end{equation}
where $M_{\rm BH}$ is the black-hole mass, and $G=6.67\times10^{-8}$\,cm$^3$\,g$^{-1}$\,s$^{-2}$ is the gravitational constant. 
For a cloud with  characteristic size $s$,
we can write:
\begin{equation}
\label{r_kepler}
 R = GM_{\rm BH}\frac{\Delta t^2}{s^2},
\end{equation}
where ${\Delta t}$ is the time difference between the start and the end of occultation.
Assuming a spherical symmetry and homogeneity, we can estimate the size of the cloud from 
the measured column density $N_{\rm H}$:
\begin{equation}
 s \approx \frac{N_{\rm H}}{n},
\end{equation}
where $n$ is the particle density.
We can now rewrite the eq.~\ref{r_kepler} as:
\begin{equation}
\label{r_kepler2}
 R \approx GM_{\rm BH}\frac{\Delta t^2 n^2}{\Delta N_{\rm H}^2},
\end{equation}
where $\Delta N_{\rm H}$ is the change
of the column density in the time interval $\Delta t$. 
We may rewrite the eq.~\ref{r_kepler2} to have a form more appropriate 
to the typical values that are measured.
We obtain an equation analogous to the eq.~3 in \citet{2002ApJ...571..234R}:
\begin{equation}
\label{r_kepler3}
\begin{split}
 R & \approx 1.335 \times 10^{13} \frac{M_{\rm BH}}{10^7M_\odot}\left(\frac{n}{10^6 {\rm cm}^{-3}}\right)^2 \\
 & \left(\frac{\Delta t}{1\,{\rm Ms}}\right)^2 \left(\frac{\Delta N_H}{10^{22}\,{\rm cm}^{-2}}\right)^{-2}  {\rm cm}.
\end{split}
\end{equation}

Using this equation, we can constrain the distance of the absorber
from the change of the column density in a given time
if we know the mass and the volume density.
The mass of the black hole is usually constrained from
different methods including reverberation mapping \citep{Peterson2014},
stellar velocity dispersions \citep{Ferrarese2000},
bulge luminosities \citep{1995ARA&A..33..581K, Magorrian1998},
or from single optical emission lines 
due to a tight correlation between the luminosity and mass \citep[see, e.g.,][and references therein]{Shen2008}.
However, the volume density is unknown. 
The density should not exceed $10^6$ particles per cm$^3$ if it is part
of a torus and it should be about $10^9$ particles per cm$^3$ if the cloud belongs to the
Broad Line Region \citep{Weedman1977}.

The volume density can be estimated if the ionisation $\xi$ of the absorber is measured:
\begin{equation}
\label{ionisation}
 n = \frac{L}{\xi R^2},
\end{equation}
where $L$ is the illuminating luminosity the energy range 13.6\,eV - 13.6\,keV.
We can now replace the density in eq.~\ref{r_kepler2} and get:
\begin{equation}
 R \approx \left[GM_{\rm BH}\frac{\Delta t^2 L^2}{\Delta N_{\rm H}^2\xi^2}\right]^{\frac{1}{5}}.
\end{equation}
Again, we can rewrite this equation with the quantities
normalised to their typical values
and we get an equation analogous to the eq.~3 in \citet{lamer2003}:
\begin{equation}
\label{r_final}
\begin{split}
R & \approx 2.66 \times 10^{17} \, \biggl[\frac{M_{\rm BH}}{10^7M_\odot}\left(\frac{\Delta t}{1\,{\rm Ms}}\right)^2\left(\frac{\Delta N_H}{10^{22}\,{\rm cm}^{-2}}\right)^{-2}\\
 & \left(\frac{L}{10^{43}\,{\rm erg\,s}^{-1}}\right)^2 \left(\frac{\xi}{\rm erg\,cm^{-2}\,s^{-1}}\right)^{-2}\biggr]^{\frac{1}{5}} {\rm cm}.
\end{split}
\end{equation}

The black-hole mass of {\source} was estimated as $M \approx 10^8 M_{\odot}$ by \citet{Bennert2006}
based on the relation between the luminosity at 5100\,{\AA}
and the black hole mass \citep{Peterson2004}.
We measured the change of the column density of the cold absorber
$\Delta N_{\rm H} \approx 4 \times 10^{22}$\,cm$^{-2}$ to be $\approx 0.45$\,Ms (see Fig.~\ref{fig_timing}).
The ionising luminosity is taken as an average value
between the 3rd and 4th observation, $L \approx 10^{44}$\,erg\,s$^{-1}$
(see Table~\ref{table_luminosities}),
and the ionisation is $\xi \approx 15$\,erg\,cm$^{-2}$\,s$^{-1}$ (see Table~\ref{model}).
After inserting these values into the eq.~\ref{r_final} we
get the estimation for the distance of the absorbing cloud from the centre:
\begin{equation*}
\label{R}
 R \approx 1.5 \times 10^{17} {\rm cm} \approx 0.05 \, {\rm pc} \approx 60\,{\rm light}\,{\rm days.} 
\end{equation*}

The distance of the Broad Line Region can be estimated from the optical luminosity.
A tight relation between the BLR radius and the luminosity at 5100$\AA$ was constrained by \citet{Kaspi2005}
and recently updated by \citet{Bentz2013} who also subtracted contributions of the host galaxies.
The luminosity of {\source} is 
$L_{\rm 5100\AA} \approx 10^{43}$\,erg\,s$^{-1}$
\citep{Bennert2006}.
This corresponds to the BLR size of $\approx 40$ light days according to \citet{Bentz2013}.
This value is consistent what we found for the variable X-ray absorber,
taking into account the general uncertainties of both estimations.

We can also derive the density of the absorbing cloud from our measurements. 
Using eq.~\ref{ionisation}, we obtain:
\begin{equation*}
 n \approx 3\times10^8 \,{\rm cm}^{-3}.
\end{equation*}
This value is slightly lower
than the averaged expected density of BLR clouds \citep{Peterson2014}.
Our measurements of slightly lower density, larger radius and low ionisation
are consistent with the reverberation in optical spectral lines 
that revealed a stratification in the ionisation and the density 
of the BLR region \citep[see, e.g.,][]{Gaskell1986}, both decreasing with the radius.

Although our measured location of the variable
absorber is fully consistent with the Broad Line Region,
we note that the innermost
part of the torus cannot be ruled out either,
as its value is not well established.
The most recent mid-infrared observations
of nearby galaxies revealed 
that there is a large scatter in the geometrical
properties of their dusty tori \citep{Burtscher2013},
and even very complicated structures of the dust
within the central parsec
were reported \citep{Tristram2014}.

\subsection{Scattering region}

The presence of a scattering region
that is not located on the line-of-sight
is suggested from the optical polarisation measurements.
Our X-ray spectra show significant spectral variability
in the energy range 2--6\,keV (see Fig.~\ref{spectrum})
that is best explained by a variable absorption. 
Less variability is evident around 1\,keV, 
which can be well explained by
a scattered power-law emission 
that is not affected by a variable absorber. 
Although its intrinsic normalisation
is lower than the normalisation of the direct power-law emission
by two orders of magnitude,
it dominates the spectrum at $\approx 1$\,keV where the direct nuclear emission 
is heavily absorbed (see Fig.~\ref{plot_model}).
%


The scattering region could be associated with the extended 
Narrow Line Region located at a rotational axis
with the similar geometry as considered by \citet{Miniutti2014} 
for another polar-scattered Seyfert~1 galaxy ESO 323-G77 
 (see their Figure~8),
or in obscured Seyfert type~2 galaxies \citep[see, e.g.,][]{bianchi2006}. 
This region is also supposed to be 
responsible for the measured polarization degree that is unusually high
for a type~1 object \citep{Schmid2001}.

An alternative explanation for the high polarisation degree 
in polar-scattered Seyfert~1 galaxies might be due to scattering
on disc winds (on the external layers of the absorbing clouds)
at the opposite side from the observer \citep{marin2013}.
\citet{Marin2014} showed on an AGN sample including {\source} 
that scattering on the disc wind better corresponds to the data 
than scattering in a polar region.
Nevertheless, to draw a clear conclusion about the geometry 
of the scattering region, 
X-ray polarimetric measurements would be required.

We note that two components were also detected for the reflection.
A part of reflected emission is coming from the innermost accretion disc
and is also affected by the absorber. The other part originates at a farther region
and is not smeared by the relativistic effects. 
The distant reflection may occur at the same 
scattering region as for the power law.

\subsection{Geometrical constraints from the reflection models}

\begin{figure}[tb]
\includegraphics[width=0.49\textwidth]{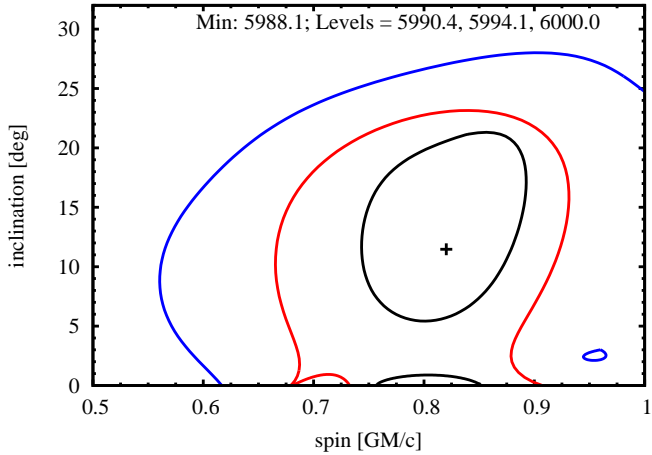}
\caption{Contour plots between the spin and the inclination
in the relativistic model. The contours show 1$\sigma$,
 2$\sigma$, and 3$\sigma$ levels.
}
\label{conia}	
\end{figure}

The reflection features have been detected mainly
owing to the broad-band coverage of the {\suzaku} satellite.
Figure~\ref{spectrum} shows the presence of both the iron line and the Compton hump. 
Although the very simple power-law model revealed
a skewed profile of the iron line especially for the 4th observation 
(see the right panel of Fig.~\ref{spectrum}),
these data residuals could be reasonably well modelled by a complex absorption
and distant reflection (see Fig.~\ref{pldelc}). 

However, the distant reflection was not able to reproduce the data residuals
at energies around 15\,keV.
We found that the Compton hump is significantly red-shifted,
which can be well accounted by relativistic reflection.
Although hard excesses have
been successfully explained also by other models,
including absorption scenario \citep[e.g.,][]{2008A&A...483..437M, Tatum2013}
or complex reprocessing in the torus \citep{Murphy2009},
we have not succeeded with simple tests using these alternatives 
to describe the data. We leave a more complex analysis using these alternatives
for future work.


Using the relativistic reflection model, we were able to measure the spin and the inclination
angle of the accretion disc,  
though with a large uncertainty given the complexity of the final model.
We obtained  $a \approx 0.8$, and $i \lesssim 20^\circ$. 
The contour plot between these two parameters is shown in Fig.~\ref{conia}
(the model parameters that are
shown in Table~\ref{model} were all left free during this contour calculations). 
The spin is not constrained sharply. Its value can be any between 0.6
and the maximum value within the 3$\sigma$ level. The inclination is lower
than it would be expected from the 
polarisation measurements that suggest the inclination to be around 45 degrees.
A possible explanation could be that the relativistic reflection 
also comprises a direct and scattered component.
The scattered part can be relatively enhanced
if the direct one is heavily obscured.
However, in the final model, the scattered component of the primary power law
is two orders of magnitudes lower than the direct part.
Assuming that the primary and disc reflection emission do not originate
far from each other, the explanation by reflection dominated by a scattered component
seems highly unlikely.

Another explanation might be 
that the innermost accretion disc has different orientation than the larger structures
\citep[see, e.g.,][]{Nayakshin2005, Ulubay-Siddiki2009, Tremaine2014}.
We can estimate a warped radius to which the accretion disc
is forced to be in the equatorial plane using a relation by \citet{Tremaine2014},
which gives $6 \times 10^{15}$\,cm for the mass $M \approx 10^8 M_{\odot}$.
This is much closer than
the distance measured for the X-ray absorbing gas ($\approx \times 10^{17}$\,cm).
This makes the possibility that the innermost accretion disc
has different orientation than the Broad Line Region.
However, a detailed analysis of the optical broad spectral lines
would be needed for a more relevant testing of this hypothesis
since warping would have an impact on the asymmetry
of the broad-line profiles \citep[see, e.g.,][]{Wu2008}.

\subsection{Origin of the Broad-Line-Region clouds}

Our detection of an absorbing gas located at the same region
as BLR may shed light on the origin and geometrical configuration of BLR.
In general, the broad optical lines are usually seen in AGNs viewed at
low inclination with a clear view of the nucleus.
This implies the BLR location close to the equatorial plane.
However, the origin and exact geometry of BLR is still rather unknown and highly debated
in the literature \citep[see, e.g.,][and references therein]{Sulentic2000, Gaskell2009}.

\citet{Elitzur2014} suggested that the BLR and torus are 
most naturally a continuation of the accretion disc outflow.
\citet{Miniutti2014} also argued that the BLR clouds might be the cold and dense
clumps of the smoother warm/hot outflow.
In {\source}, we have also detected highly ionised absorbers with an out-flowing velocity comparable 
to the velocity derived from the width of the optical spectral lines.
However, the location of the highly ionised warm absorber 
cannot be directly constrained from our data since 
there is no strong evidence for its variability. 
If it is part of BLR, it must constitute a large volume
and have a very low density. However, it can also originate 
closer to a black hole and be part of the innermost outflow. 
The higher ionisation can be then explained by a stronger 
illumination from the centre.

\citet{Czerny2011} proposed another origin of the broad-line clouds.
They found that the temperature at BLR
is close to the dust sublimation temperature, which gives to arise dusty winds
from the accretion disc. Once they get up from the equatorial plane 
the strong irradiation destroys the dust.
The BLR clouds lose support from the radiation pressure,
which is large on dusty particles but much smaller on pure-gas clouds.
As a result, the clouds are falling back to the disc
and again arise.
This process could explain why the BLR clouds get to the line-of-sight 
of the intermediate type Seyferts like {\source}.

Moreover, the presence of the ionised outflow can provide
a larger protection of the dusty winds from the destroying irradiation
and the clouds may get farther
from the equatorial plane.
Large anti-correlation of the column density of the absorber 
with the intrinsic flux of the nucleus (see Fig.~\ref{rel_intflux_abs})
is consistent with this scenario.
Lower flux would allow dusty winds to arise higher
implying that more clouds may enter into the line-of-sight,
and thus increase the observed column density of the cold absorber.
This relation can, however, be also
caused by the changes in the gas opacity due to
the decrease of illumination \citep[see, e.g.,][]{Krongold2005}.
However, with no significant measurements of the change in the ionisation of the absorber,
we have no quantitative indication for the temperature changes
that would cause the changes in the gas opacity.

We note that our location estimation is based on assuming the
Keplerian orbital velocity. 
\citet{Czerny2015} showed that the vertical velocity
of the clouds is of order of $z/R \times \Omega_{\rm K}$.
The $z/R$ fraction should correspond
to the fraction of sky that is covered by the BLR clouds,
which is from the observations of the order of $\lesssim 0.3$ \citep[see, e.g.,][]{goad1998}.
This satisfies the usage of the Keplerian velocity 
as the estimated velocity of the absorbing cloud.

Our results also indicate that the distance of the X-ray absorbing cloud(s)
is beyond the warped radius. The warped radius
may represent an outer radius of the accretion disc,
because the disc becomes unstable beyond it \citep[see, e.g.,][]{collin2007}. 
This allows a possibility that the BLR clouds
originate due to a fragmentation of still unstable accretion flow.
Such clouds would not feel strong gravitational attraction
towards the disc plane and may reach a more inclined orbit.
This could be an alternative explanation of the origin of the BLR clouds
and why they may get into the line-of-sight.

\section{Conclusions}
\label{conclusion}

We performed a monitoring programme of {\source}
to constrain the characteristic time scale of its X-ray 
spectral variability using the {\suzaku} satellite.
We obtained four new observations separated by 1.5, 2 and 5.5 day intervals.
A significant spectral change was detected only for the last observation
with the half flux compared to the first observation,
although the flux also varied during the individual observations.
By comparison of the light curves in the X-ray soft (0.5--10\,keV) 
and hard (15--55\,keV) energy bands, 
we found that the variability occurs almost entirely at the soft X-rays.
The 5-days long variability implies the location of the absorber
to be most likely in the Broad Line Region.

Our analysis revealed the spectral complexity in the X-ray band. 
The primary nuclear emission can be well described
by a power law with the photon index $\Gamma \approx 2$.
It is strongly affected by at least three absorbers
with different ionisations ($\log\xi \approx$ 1 -- 4, respectively),
the lower ionised one(s) being variable and 
causing the spectral variability. 
The spectral variability is most prominent at 2--6\,keV,
while the part of the spectrum at $\lesssim$ 1\,keV is probably 
dominated by a scattered power-law emission 
coming from either a polar region or an external part of the ionised wind itself.

The hard X-ray spectrum revealed an excess at $\approx 15$\,keV,
which can be well explained by a reflection from an accretion disc
that is affected by the relativistic smearing
due to strong gravitational field around the super-massive
black hole and very high orbital velocities in the accretion disc.
Using the relativistic reflection model,
we measured the black hole spin, $a \approx 0.8$,
but with large uncertainty due to the model complexity.

The spectral complexity and the rapid absorption variability
makes {\source} a very prominent source for further studies
with X-ray instruments.
In particular, detectors with high-quality spectral resolution
would help to better constrain the ionisation structure
of the absorber and its response to the continuum
flux changes. Future high-resolution observations of {\source} with,
e.g., planned X-ray mission Astro-H will be therefore desirable.



\section{Acknowledgements}

The authors acknowledge suggestions 
by an anonymous referee who significantly helped to
improve the paper.
JS also acknowledges useful discussions with Leonard Burtscher,
Giovanni Miniutti, Bozena Czerny and Frederic Marin.
This research was financially supported from the Grant Agency of the Czech Republic
within the project No.\,14-20970P.
The research was also partially funded by 
the European Union Seventh Framework Program (FP7/2007–2013) under grant 312789
and the Bundesministerium
f\"ur Wirtschaft und Technologie through Deutsches Zentrum f\"ur Luft-
und Raumfahrt Grant 50OR1311.

\bibliographystyle{aa} 

\bibliography{/home/jirka/Documents/references} 




\end{document}